\documentclass[10pt,aps,prl,floatfix,showpacs,twocolumn,superscriptaddress]{revtex4-2}

\usepackage{graphicx}
\usepackage{amsmath, amsfonts, amssymb, bm,color}

\usepackage{amstext}
\usepackage{mathrsfs}
\usepackage{pdfpages}
\usepackage{pgffor}
\usepackage{etoolbox} 

\makeatletter
\patchcmd{\@outputpage@head}{\@ifx{\LS@rot\@undefined}{}{\LS@rot}}{}{}{}
\makeatother

\begin{document}

\title{Radiation Reaction Enhancement in Flying Focus Pulses}
\author{M. Formanek}
\email{formanek@mpi-hd.mpg.de}
\affiliation{Max Planck Institute for Nuclear Physics, Saupfercheckweg 1, D-69117 Heidelberg, Germany} 
\author{D. Ramsey}
\author{J. P. Palastro}
\affiliation{University of Rochester, Laboratory for Laser Energetics, Rochester, New York, 14623 USA}
\author{A. Di Piazza}
\email{dipiazza@mpi-hd.mpg.de}
\affiliation{Max Planck Institute for Nuclear Physics, Saupfercheckweg 1, D-69117 Heidelberg, Germany} 
\begin{abstract}
Radiation reaction (RR) is the oldest still-unsolved problem in electrodynamics. In addition to conceptual difficulties in its theoretical formulation, the requirement of exceedingly large charge accelerations has thus far prevented its unambiguous experimental identification. Here, we show how measurable RR effects in a laser-electron interaction can be achieved through the use of flying focus pulses (FFPs). By allowing the focus to counterpropagate with respect to the pulse phase velocity, a FFP overcomes the intrinsic limitation of a conventional laser Gaussian pulse (GP) that limits its focus to a Rayleigh range. For an electron initially also counterpropagating with respect to the pulse phase velocity, an extended interaction length with the laser peak intensity is achieved in a FFP. As a result, the same RR deceleration factors are obtained, but at FFP laser powers orders of magnitude lower than for ultrashort GPs with the same energy. This renders the proposed setup much more stable than those using GPs and allows for more accurate \emph{in situ} diagnostics. Using the Landau-Lifshitz equation of motion, we show numerically and analytically that the capability of emerging laser systems to deliver focused FFPs will allow for a clear experimental identification of RR.
\end{abstract}

\maketitle
Radiation reaction (RR), i.e., the energy and momentum loss of an accelerated charge as it emits radiation, remains an outstanding issue in the formulation of classical electrodynamics \cite{Landau_b_2_1975,Barut_b_1980,Rohrlich_b_2007}. The classical equation of motion accounting for RR, the Lorentz-Abraham-Dirac equation (LAD) \cite{Dirac_1938}, suffers from causality issues, runaway solutions, and/or problems with initial conditions. The Landau-Lifshitz (LL) equation \cite{Landau_b_2_1975} is free from these shortcomings, but it is derived from the LAD equation. Thus, experimentally testing the classical RR equation is still an outstanding and important problem. Alternative classical RR equations, such as the Eliezer-Ford-O'Connell equation, are indistinguishable at the classical level from the LL equation, because they differ by terms smaller than quantum corrections \cite{Koga_2004,Hadad_2010,Bulanov_2011}. To this day, RR remains an active area of investigation highlighted by a number of research \cite{Vranic_2014,Blackburn_2014,Tamburini_2014,Li_2014,Heinzl_2015,Yoffe_2015,Capdessus_2015,Vranic_2016,Dinu_2016,Di_Piazza_2017,Harvey_2017,Ridgers_2017,Niel_2018a,Niel_2018b,Formanek_2020,Di_Piazza_2021b} and review articles published over the last decade \cite{Hammond_2010,Di_Piazza_2012,Burton_2014,Blackburn_2020,Gonoskov_2021}, as well as in recent experimental efforts to measure the effects of RR on electrons interacting with aligned crystals \cite{Wistisen_2018,Nielsen_2021} and ultrastrong laser fields \cite{Cole_2018,Poder_2018}. Apart from its fundamental importance, relating, e.g., to intrinsic properties of elementary particles like the mass of the electron, RR plays a crucial role in several fields of physics, such as astrophysics, plasma, and accelerator physics. 
 
Progress in RR research is mainly hindered by the experimental difficulty of its detection. A number of experimental facilities, including synchrotrons, wigglers, and x-ray free electron lasers, employ an external electromagnetic field to wiggle an electron and produce radiation. However, because the emitted energy is much smaller than the electron energy, even when accounting for electron beam coherence effects, the effect of RR on the electron trajectory is negligible. Furthermore, recent experiments utilizing high-intensity lasers \cite{Cole_2018,Poder_2018} operated in a regime where quantum effects ``interfered'' with classical RR, complicating their physical interpretation.

The flying focus is a newly developed technique for controlling the trajectory of peak laser intensity over distances much longer than the Rayleigh range \cite{Sainte-Marie_2017,Froula_2018}. In the original experimental demonstrations, the peak intensity was made to travel at any desired velocity by adjusting the chirp and using a chromatic lens to independently set the time and location at which each frequency within the pulse came to focus \cite{Sainte-Marie_2017,Froula_2018}. More recent implementations have proposed axiparabola-echelon optics \cite{Palastro_2020} and \lq space-time light sheets\rq\ \cite{Kondakci_2017,Yessenov_2020} to achieve the same effect. Building on this capability, several studies have illustrated the advantage of flying focus pulses (FFPs) for a wide range of laser-based applications, including ionization waves in plasma \cite{Turnbull_2018, Palastro_2018}, photon acceleration \cite{Howard_2019}, laser wakefield acceleration \cite{Palastro_2020}, vacuum electron acceleration \cite{Ramsey_2020}, and nonlinear Thomson scattering \cite{Ramsey_2021}.

\begin{figure}
	\begin{center}
		\includegraphics[width=\columnwidth]{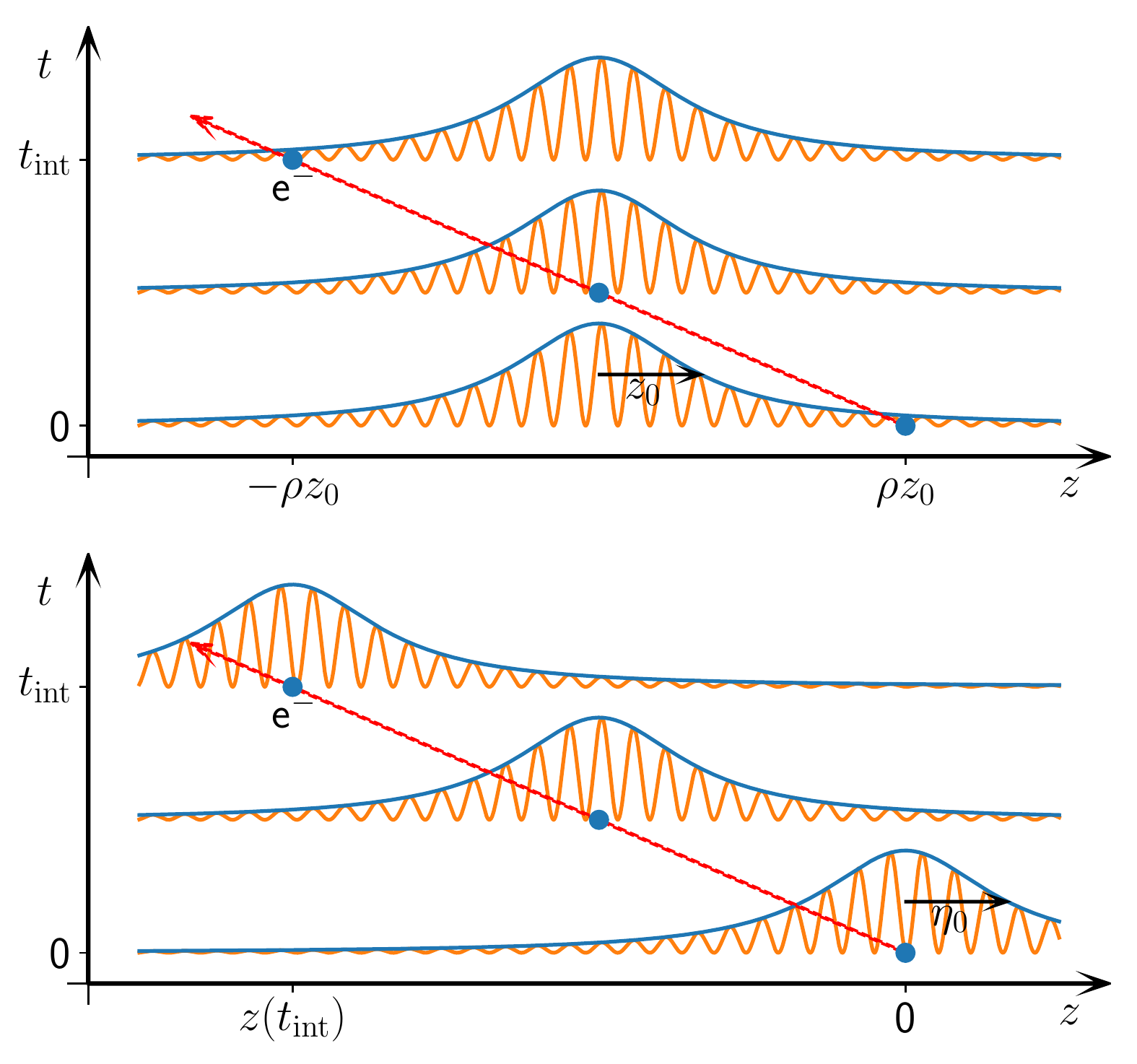}
		\caption{\label{fig:beam_illustration} Schematic representation of an ultrarelativistic electron counterpropagating with respect to a Gaussian beam (top panel) and to a flying focus beam with focal velocity equal and opposite to the phase velocity (bottom panel). For the sake of clarity the laser-pulse envelope is not included (see the text and SM \cite{Supplemental} for details). The axes are not to scale.}
	\end{center}
\end{figure}

In the present Letter we show that FFPs lower the laser power required for significant RR deceleration of electrons (charge $e<0$ and mass $m$) by orders of magnitude compared to conventional ultrashort Gaussian pulses (GPs). The high-intensity region of a GP is set to the Rayleigh range which defines a limited spatial domain through which an ultrarelativistic electron quickly passes. This is especially true for ultrashort GPs, which have their pulse energy concentrated to a fraction of the Rayleigh range. In contrast, the peak intensity of a FFP can move at the speed of light and in the opposite direction of its laser phase velocity (Fig. \ref{fig:beam_illustration}). Thus, an ultrarelativistic electron traveling in the opposite direction of the phase fronts can remain in the \lq focus\rq\ of a FFP for extended interaction times limited only by the total pulse energy. In order to clearly compare the performances of both field configurations, we first analytically calculate the electron energy loss. Then, we validate the FFP results numerically by simulating the electron trajectories using the LL equation for RR \cite{Landau_b_2_1975}. The lower power and peak intensity required by FFPs minimize the quantum effects, provide additional control, and improve diagnostic access to unambiguously identify this elusive phenomenon in experiments.

It was shown in Ref.~\cite{Di_Piazza_2021} that the exact solution of Maxwell's equations given in Ref.~\cite{Esarey_1995} describes a monochromatic flying focus beam (FFB) with a fixed focal velocity $v_f = -1=-v_p$, with $v_p$ being the beam phase velocity (units with $\hbar=c=\epsilon_0=1$ are used throughout). Here, we refer to ``beams'' (GBs/FFBs) in the infinite, monochromatic case and to ``pulses'' (GPs/FFPs) in the finite, time-localized case. We employ this solution to model the FFBs because (i) it satisfies the vacuum wave equation exactly; (ii) the electric and magnetic fields can be expressed analytically in closed form; (iii) its exponential drop-off in the transverse direction assures a finite beam power, which is important for a direct comparison with GBs. We indicate as $A^\mu(x)$ the four-vector potential of either the FFB or the GB and we work within the Lorenz gauge $\partial \cdot A = 0$. In the FFB case we impose the additional condition $A_+(x)=A^0(x) + A^z(x) = 0$ \cite{Di_Piazza_2021}. We consider an expression of the four-vector potential, which is an exact solution of the vacuum wave equation $\partial^2 A^\mu = 0$, in the case of a monochromatic spectral profile (see Ref.~\cite{Di_Piazza_2021} for the case of the Gaussian spectral profile and the Supplemental Material (SM) \cite{Supplemental}). For a FFB polarized along the $x$-axis with a wave-vector pointing in the direction of the positive $z$-axis, the independent four-potential components are
\begin{subequations}
\begin{align}
\label{eq:FFB1}A^x &= \mathcal{A}_0 \frac{\sigma_0}{\sigma(\eta,\eta_0)} e^{-r^2/\sigma^2(\eta,\eta_0)}\cos [\Psi(0,\eta,\eta_0)]\,,\\
\label{eq:FFB2}A^0 &= \frac{\mathcal{A}_0}{\omega_0} \frac{x}{\sigma^2(\eta,\eta_0)} e^{-r^2/\sigma^2(\eta,\eta_0)}\sin [\Psi(1,\eta,\eta_0)]\,.
\end{align}
\end{subequations}
Here, we have introduced the four-potential amplitude $\mathcal{A}_0$, the spot radius $\sigma_0$, the angular frequency $\omega_0 = 2\pi/\lambda_0$, and the laser wavelength $\lambda_0$ as the main quantities characterizing the beam. Also, we employ light-cone coordinates $\phi = t - z$, $\eta = t + z$, and $\bm{r}=(x,y)$, such that $r = \sqrt{x^2 + y^2}$ is the distance from the $z$-axis, and $\sigma(\eta,\eta_0) = \sigma_0 \sqrt{1+ \eta^2 / \eta^2_0}$, $\eta_0 = \omega_0 \sigma^2_0$. This implies that the focus of the FFB is placed at $\eta = t + z = 0$, i.e., the focal velocity is $-1$, opposite to the propagation direction of the phase fronts. Finally, the phase $\Psi(a,\eta,\eta_0)$ is defined as
\begin{equation}\label{eq:FFphase}
\Psi(a,\eta,\eta_0) = \omega_0 \phi - \frac{r^2}{\sigma^2(\eta,\eta_0)}\frac{\eta}{\eta_0} + (1 + a)\arctan\left(\frac{\eta}{\eta_0}\right)\,.
\end{equation}
 
For the GBs we employ the solution within the paraxial approximation in which the diffraction angle $\theta = \sigma_0 / z_0$ is the small parameter \cite{Salamin_2007}. Here, $z_0 = \omega_0 \sigma_0^2/2$ is the Rayleigh length. We again consider a linearly polarized field in the $x$-direction with the wave vector pointing along the positive $z$-axis. The solution of the paraxial equation within the Lorenz gauge and with $A^z = 0$ is given by 
\begin{subequations}
\begin{align}
\label{eq:GB1}A^x &= \mathcal{A}_0 \frac{\sigma_0}{\sigma(z,z_0)}e^{-r^2/\sigma^2(z,z_0)}\cos[\Psi(0,z,z_0)]\,,\\
\label{eq:GB2}A^0 &= \frac{\mathcal{A}_0}{\omega_0} \frac{2x}{\sigma^2(z,z_0)} e^{-r^2/\sigma^2(z,z_0)}\sin[\Psi(1,z,z_0)]\,,
\end{align}
\end{subequations}
which places the stationary focus of the GB at $z=0$. 

The time-averaged power of the GB going through the $xy$-plane can be expressed in the paraxial approximation as \cite{Esarey_1993}
\begin{equation}\label{eq:PGB}
P_\text{ave} = \frac{\pi}{4}\mathcal{A}_0^2\omega_0^2 \sigma_0^2\approx 21.5\, \text{GW} \left(\xi_0 \frac{\sigma_0}{\lambda_0}\right)^2\,,
\end{equation}
where $\xi_0=|e|\mathcal{A}_0 / m$ is the dimensionless normalized amplitude, which is related to the laser peak intensity $I_0$ as $I_0(\text{W}/\text{cm}^2) = 1.37\times 10^{18} \xi_0^2 (\lambda_0[\mu\text{m}])^{-2}$. The corresponding expression for the FFB is the same (see the SM \cite{Supplemental}). The time-averaged power in both cases is derived under the assumption that the Rayleigh length is much larger than the laser wavelength.

In order to transition from monochromatic beams to {\it pulses} of finite energy, we employ a slowly varying envelope $g(\phi)$ with a constant flat-top profile (see SM \cite{Supplemental}). We work in an approximation of long pulses and neglect any derivatives of the envelope $g(\phi)$. For a total pulse energy $E_\text{tot}$ and average power $P_\text{ave}$, the pulse length is given by $\tau = E_\text{tot} / P_\text{ave}$. If spatial focusing effects are ignored, i.e., for a plane wave characterized by the envelope $g(\phi)$, and if a pulse counterpropagating with respect to an ultrarelativistic electron is considered, then the wave-electron interaction time $t_\text{int}$ is approximately given by $\tau/2$.

Since we are going to consider ultrarelativistic electrons at the focus of the laser field, for the sake of analytical estimations, we assume that the latter can be locally approximated as a plane wave with the dimensionless amplitude $\xi(t)$ given by the field value at $r=0$. Also, in the ultrarelativistic limit the electron energy loss can be directly computed from the relativistic Larmor formula $P_\text{L} = -(2/3)m r_e \dot{u}^2$ of the electromagnetic radiated power [we use the diagonal metric tensor $(+1,-1,-1,-1)$]. Here, $r_e = e^2 / (4\pi m)$ is the classical electron radius and $\dot{u}^\mu$ is the proper-time derivative of the four-velocity $u^{\mu}=(\gamma,\pmb{u})$. This corresponds to the energy loss $d\gamma/dt = (2/3)r_e \dot{u}^2$, where $\dot{u}^2 = - \xi^2(t) (k_0\cdot u)^2$ in a plane-wave with four-wave-vector $k_0^\mu = (\omega_0,\bm{k}_0)$. For an ultrarelativistic electron moving in the direction opposite the wave vector $\bm{k}_0$, $k_0\cdot u\approx 2\omega_0 \gamma$ and $\phi\approx 2t$ along the electron trajectory. Thus, the differential equation for the electron gamma factor $\gamma(t)$ with the initial condition $\gamma(0)=\gamma_0$ has the approximate solution \cite{Di_Piazza_2008_a}
\begin{equation}\label{eq:gammat}
\gamma(t) \approx \frac{\gamma_0}{1 + \kappa(t)}\,,
\end{equation}
where $\kappa(t)=\frac{4}{3}\gamma_0 r_e \omega_0^2\int_0^t g^2(t')\xi^2(t')dt'$ represents the deceleration factor after a time $t$ and where the integral is taken over the slowly varying amplitude function $\xi(t)$ and envelope $g(t)$ to be computed along the electron trajectory at $r=0$. For the analytical estimates, we assume a unit rectangular envelope $g(t) = 1$ for $t \in (0, t_\text{int})$ and zero elsewhere.

In the GP case the amplitude changes as $\xi(t) = \xi_0 / \sqrt{1+z^2(t)/z_0^2}$. We assume the best-case scenario where the electron interacts with the pulse while moving through the region of its highest focus. Thus, the electron trajectory is approximately given by $r(t)=0$ and $z(t) = \rho z_0 - t$, where $r(0)=0$ and $z(0) = \rho z_0$ is the initial electron position, with $\rho$ being a dimensionless parameter defined according to the following considerations. We set the ``final'' electron position at $t=t_\text{int}$, i.e., after moving through the whole focal region at almost the speed of light, to the value $z(t_\text{int}) = -\rho z_0$ (see the top panel of Fig. \ref{fig:beam_illustration}). Thus $t_\text{int} = 2\rho z_0$, and the parameter $\rho$ gives half of the number of Rayleigh lengths $z_0$ over which the electron interacts with a GP with fixed total pulse energy $E_\text{tot}$ and average power $P_\text{ave}$: $\rho = E_\text{tot}/(4z_0 P_\text{ave})$. The integral for $\kappa(t)$ can be evaluated as $\int_0^{t_\text{int}} \xi^2(t')dt' =  2\xi_0^2 z_0 \arctan(\rho)$. By using Eq. (\ref{eq:PGB}) for the average power $P_\text{ave}$, the deceleration factor $\kappa_\text{GP}$ after the interaction time $t_\text{int}$ can be expressed as
\begin{equation}\label{eq:kappaGP}
\begin{split}	
\kappa_\text{GP}&(t_\text{int}) = \frac{32}{3}\frac{E_\text{tot} \mathcal{E}_0}{m^2} \left(\frac{r_e}{\sigma_0} \right)^{2} \frac{\arctan(\rho)}{\rho}\\
&\approx 2.0\frac{E_\text{tot}[\text{J}]\ \mathcal{E}_0(\text{GeV})}{\sigma_0^2[\mu\text{m}]}\ \frac{\arctan(\rho)}{\rho}\,,
\end{split}
\end{equation}
where $\mathcal{E}_0 = m\gamma_0$ is the initial electron energy. This means that at fixed pulse energy $E_\text{tot}$ the deceleration factor is larger for smaller focal spot sizes $\sigma_0 \rightarrow 0$ and smaller interaction times $\rho \rightarrow 0$. Both of these trends require increasing the pulse amplitude and power to keep the total pulse energy $E_\text{tot}$ fixed. This can be seen from the relation $P_\text{ave} = E_\text{tot} /2 t_\text{int} \sim E_\text{tot} / \sigma_0^2 \rho$ and, taking into account Eq. (\ref{eq:PGB}), $\xi_0^2 \sim P_\text{ave}/\sigma_0^2 \sim E_\text{tot} / \sigma_0^4 \rho$. 

In principle, the deceleration factor can be arbitrarily large (until the electron stops) but, as we decrease $\sigma_0$ to about $2\lambda_0$, we run into issues with the paraxial approximation, with the assumptions for deriving Eqs. (\ref{eq:PGB}) and  (\ref{eq:gammat}), not to mention the difficulties in the experimental feasibility of such pulses \cite{yoon2021realization}. For a specified total pulse energy $E_\text{tot}$ and average power $P_\text{ave}$ the interaction parameter $\rho$ is given by
\begin{equation}\label{eq:npfix}
\rho = \frac{1}{2\omega_0 \sigma_0^2}\frac{E_\text{tot}}{P_\text{ave}} \approx \frac{2.4 \times 10^{-2} E_\text{tot}[\text{J}]\lambda_0[\mu \text{m}]}{P_\text{ave}[\text{PW}]\ \sigma_0^2[\mu\text{m}]}\,.
\end{equation}
The estimate for the deceleration factor $\kappa$ is then obtained by substituting this expression into Eq. (\ref{eq:kappaGP}). 

For the FFP the situation is considerably simpler because the electron can co-travel with the moving focus for the duration of the interaction $t_\text{int}$ (see bottom panel of Fig. \ref{fig:beam_illustration}). Then, the integrand in $\kappa(t)$ is constant and $\int_0^{t_\text{int}} \xi^2(t')dt' = \xi_0^2t_\text{int}$. By using the expression of the power $P_\text{ave}=E_\text{tot}/2t_\text{int}$ and Eq.~(\ref{eq:PGB}), we obtain the final deceleration factor 
\begin{equation}\label{eq:kappaFFP}
\begin{split}	
\kappa_\text{FFP}(t_\text{int}) &= \frac{32}{3}\frac{E_\text{tot} \mathcal{E}_0}{m^2} \left(\frac{r_e}{\sigma_0} \right)^{2}\approx 2\frac{E_\text{tot}[\text{J}] \mathcal{E}_0[\text{GeV}]}{\sigma^2_0[\mu\text{m}]}\,.
\end{split}
\end{equation}
We note that this result does not depend on $P_\text{ave}$. Thus, one can obtain the same deceleration effect by decreasing the average power, provided that the interaction time $t_\text{int} = E_\text{tot}/2P_\text{ave}$ increases accordingly. In other words, FFPs allow us to decrease the beam power in a trade-off for a longer interaction time. At fixed total energy and spot size the scaling with the interaction time is $P_\text{ave} \propto \xi_0^2 \propto t_\text{int}^{-1}$. From Eq. (\ref{eq:gammat}) we have
\begin{equation}
\xi_0^2 = \frac{3}{16\pi^2}\frac{\kappa_\text{FFP}\lambda_0^2}{t_\text{int} r_e}\frac{m}{\mathcal{E}_0} \approx 11.5 \frac{\kappa_\text{FFP}\lambda_0^2[\mu\text{m}]}{t_\text{int}[\text{ps}]\mathcal{E}_0[\text{GeV}]} \,.
\end{equation}	
Analogously, for fixed $P_\text{ave}$ the spot size can grow with interaction time as $\sigma_0 \propto \sqrt{t_\text{int}}$ while keeping the overall deceleration constant. This is not possible for GPs whose interaction with charged particles is limited by the Rayleigh length [see Eq. (\ref{eq:kappaGP})].

\begin{figure}
	\begin{center}
		\includegraphics[width=\columnwidth]{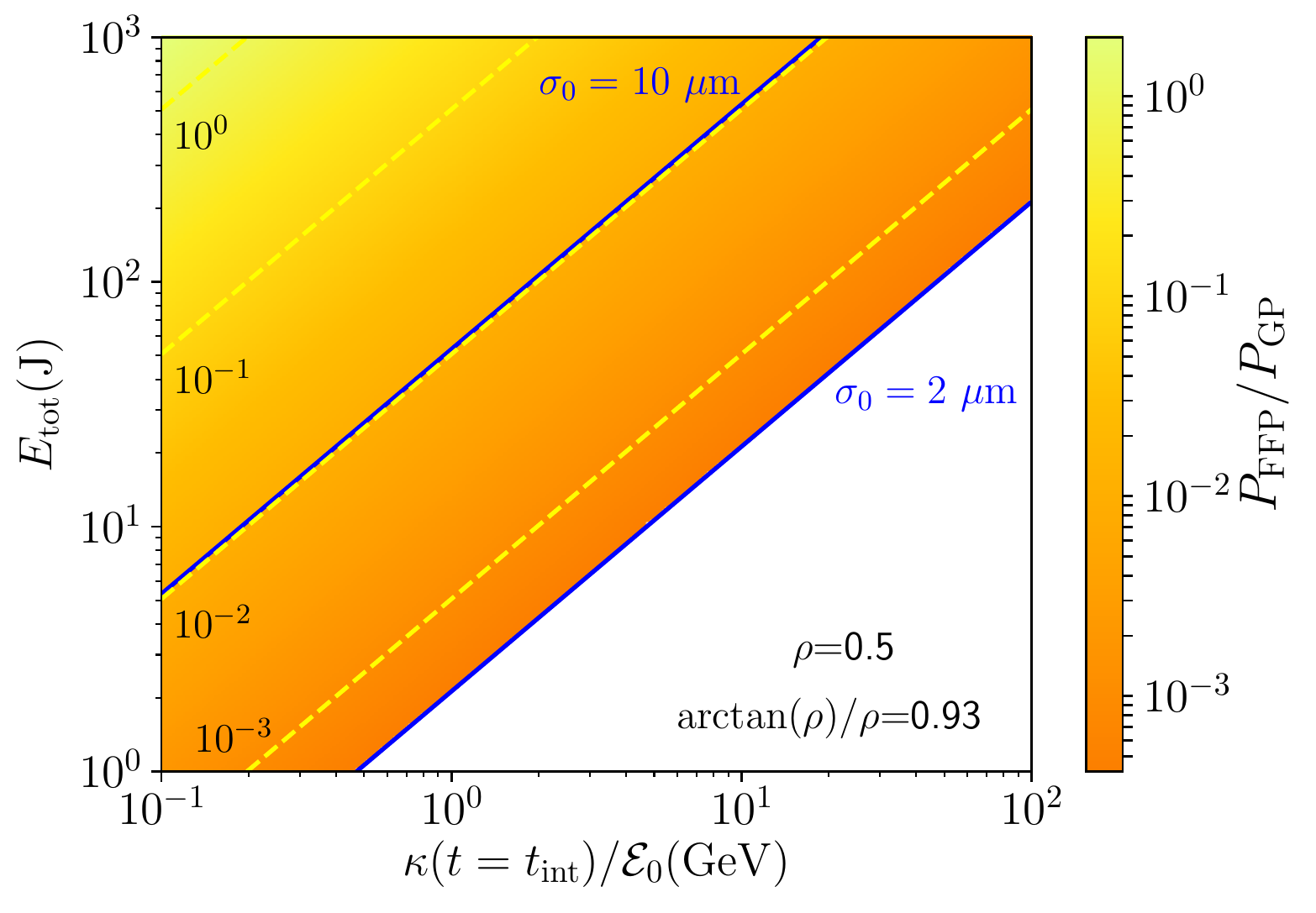}
		\caption{\label{fig:power_comparison}The necessary average power in $t_\text{int}$ = 100 ps FFPs vs the average power in GPs for a desired deceleration and a given pulse energy. The dashed yellow lines correspond to $\log_{10}(P_\text{FFP} / P_\text{GP}) \in \{-3,-2,-1,0\}$. The solid blue lines mark the range $\sigma_0 \in (2,10)\ \mu$m for GPs as indicated. The same boundaries for FFPs are with $\rho = 0.5$ almost identical. The plot is cut off at GP $\sigma_0 = 2\lambda_0=2\ \mu$m [see the discussion above Eq. (\ref{eq:npfix})].}
	\end{center}
\end{figure}

For GPs with a pulse length longer than their Rayleigh range $(\rho > 1)$, the factor $\arctan(\rho)/\rho$, by which equations (\ref{eq:kappaGP}) and (\ref{eq:kappaFFP}) differ, goes to zero. At fixed total pulse energy, electrons in FFPs achieve higher decelerations than in GPs by a factor $\rho/\arctan(\rho)>1$. In state-of-the-art high intensity laser systems, the pulses are already compressed to a very small fraction of the Rayleigh range around the focus ($\rho \rightarrow 0$) \cite{Cole_2018,Poder_2018} and at the same total energy the FFP improvement is only marginal ($\arctan(\rho)/\rho \rightarrow 1$). In this situation, however, FFPs can achieve the same deceleration for much lower laser powers by increasing $t_\text{int}$. This is crucial for precision RR experiments where the lower laser power and intensity provide better control over the interaction environment and enables \textit{in situ} diagnostics, e.g., for the laser intensity, which are unavailable at ultrahigh fields \cite{Cole_2018,Poder_2018}. 

In Fig. \ref{fig:power_comparison} we show the improvement in necessary average power in $t_\text{int}$ = 100 ps FFPs over compressed GPs with $\rho = 0.5$. For such $\rho$ the electron interacts with exactly one Rayleigh range of the GP. Although the decelerations for the same energy and spot size are almost identical in this example, FFPs can achieve the same with up to a thousand times less power. As the GPs become longer ($\rho > 1$), their power requirements also decrease, but high decelerations are no longer accessible at given energy due to the limited extent of their focal region. If we would increase the energy in the GP to keep the deceleration constant (at given $\sigma_0$) it would grow quickly with $\rho$ as $E_\text{tot} \propto \rho/\arctan(\rho)$ but the necessary power would decrease slowly as $P_\text{GP} \propto 1/\arctan(\rho)$.

In order to demonstrate the cumulative nature of RR deceleration in a FFP, we have numerically solved for the electron motion using the LL equation \cite{Landau_b_2_1975}
\begin{equation}
\dot{u}^\mu = \mathcal{F}^{\mu\nu}u_\nu + \frac{2}{3}r_e \left[\dot{\mathcal{F}}^{\mu\nu}u_\nu + (\delta^\mu_\nu - u^\mu u_\nu) \mathcal{F}^{\nu\alpha}\mathcal{F}_{\alpha \beta}u^\beta \right]\,,
\end{equation}
where $\mathcal{F}^{\mu\nu}= e(\partial^\mu A^\nu - \partial^\nu A^\mu)/m$. The first term alone (Lorentz force) would not account for particle deceleration and the electron would not undergo net energy loss. We have ensured numerically that the term proportional to $\dot{\mathcal{F}}^{\mu\nu}$ is negligible, see, e.g., also Refs. \cite{Tamburini_2010,Li_2021} and omitted it from the simulations. The focus of FFPs was successfully propagated in experiments for distances $\sim 0.5$ cm ($t_\text{int}\approx 16$ ps) \cite{Froula_2018}. In our simulations we fixed the laser wavelength at $\lambda_0 = 1\ \mu$m and interaction time  $t_\text{int} = 100\ \text{ps} \approx 1.884\times 10^{5}\ 1/\omega_0$, which can be achieved by increasing the chirp relative to a $t_\text{int}\approx 16$ ps. The total pulse energy was set to $E_\text{tot} =$ 10, 50, and 200 J, corresponding to $P_\text{ave} =$ 0.05, 0.25, and 1 TW, respectively, and to $\xi_0$ varying in the range 0.19 - 2.7 (see SM \cite{Supplemental}), i.e., peak intensities $I_0 = 5
\times 10^{16} - 1\times 10^{19}$ W/cm$^2$. The initial electron gamma factor was $\gamma_0 = 1000$ ($\mathcal{E}_0 = 0.511$ GeV) and the laboratory time step was set to $dt=0.01\ 1/\omega_0$. The quantum nonlinearity parameter $\chi_0 = 5.9 \times 10^{-2}\  \mathcal{E}_0[\text{GeV}] \sqrt{I_0[10^{20}\ \text{W/cm}^2]}$ is in the range $9.5\times 10^{-3}-6.7\times 10^{-4}$ justifying the classical treatment of RR \cite{Di_Piazza_2012}. Finally, the above-mentioned envelope function $g(\phi)$ was implemented as a smooth, symmetric, 5th-order polynomial rise and fall surrounding a constant flat-top profile. The $t_\text{int}$ = 100 ps pulse is sufficiently long that the envelope can vary slowly compared to $\eta_0$ and the pulse still maintains the approximately rectangular shape (see SM \cite{Supplemental}).

\begin{figure}
	\begin{center}
		\includegraphics[width=\columnwidth]{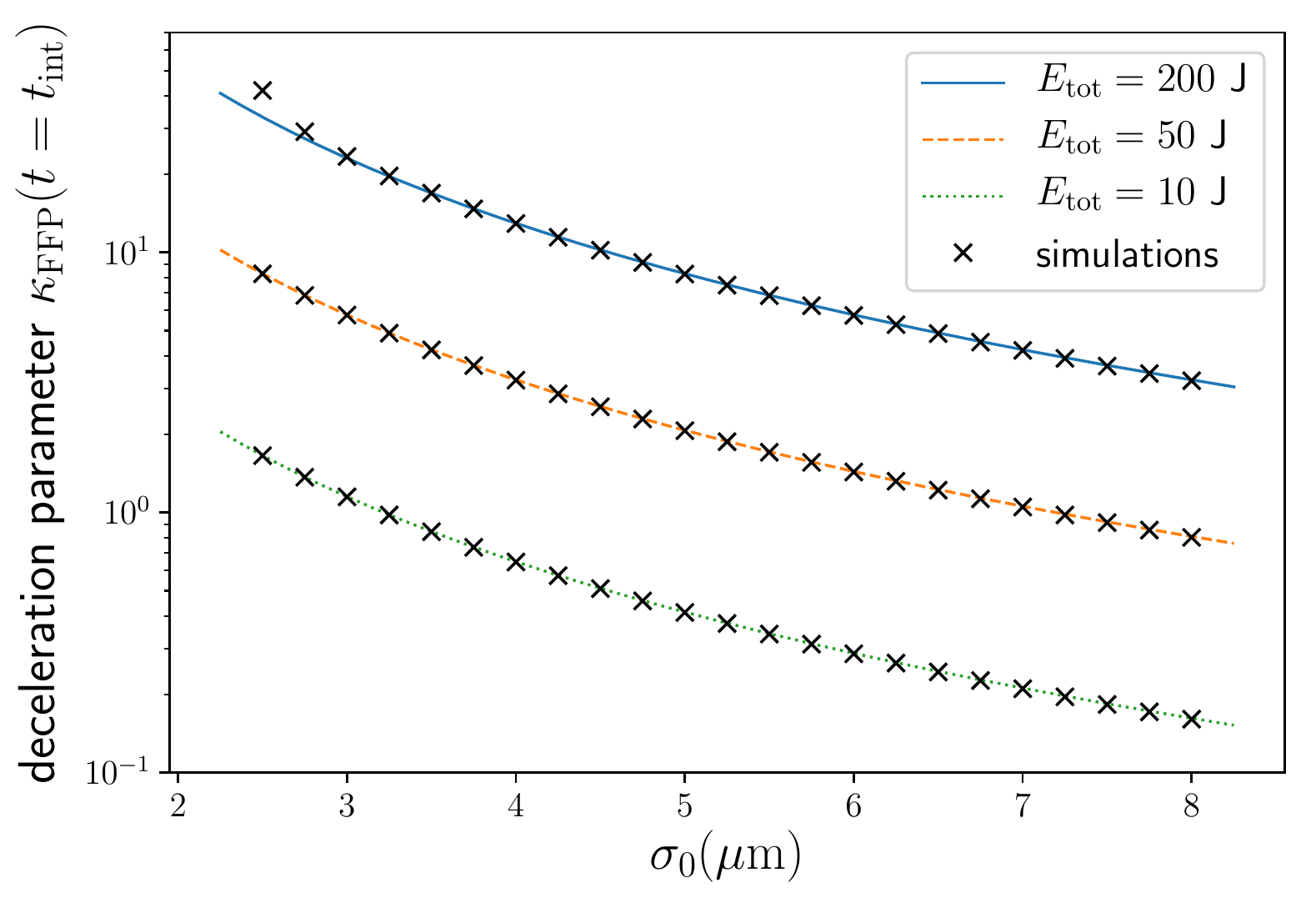}
		\caption{\label{fig:numerical_test} Overall RR deceleration $\kappa_\text{FFP}$ after interaction time $t_\text{int} = 100$ ps of an electron with $\mathcal{E}_0 = 0.511$ GeV in the FFPs (see the text for other numerical parameters). The curves are analytical estimates and the crosses result from the numerical simulations.}
	\end{center}
\end{figure}

Figure \ref{fig:numerical_test} demonstrates that the FFP energy loss estimates from Eq. (\ref{eq:kappaFFP}) are in an excellent agreement with the numerical results except for the highest $\kappa_\text{FFP}(t_\text{int})$. Once the electron is decelerated to $\gamma \lesssim 30$, the interaction with the pulse becomes more complicated than our estimates capture. For example, the electron begins to lag behind the FFP and experiences additional ponderomotive deceleration \cite{Ramsey_2021}. Further, the transverse oscillations in the field become important and the approximation $k_0 \cdot u \approx 2\omega_0\gamma$ used for deriving Eq. (\ref{eq:gammat}) is no longer valid.

In conclusion, we have shown that FFPs allow one to reach significant RR deceleration effects with orders of magnitude lower laser power than ultrashort Gaussian pulses currently used in experimental attempts to measure RR. This was achieved by exploiting the cumulative nature of RR effects and the unique properties of the FFPs, for which the peak intensity can move in the opposite direction of the phase velocity. In contrast to GPs, which require a high degree of temporal compression to reach the necessary intensity, a long FFP pulse can be used, alleviating technological constraints on the optics \cite{stuart1995laser} and allowing for \emph{in situ} diagnostics.

Previous experiments \cite{Froula_2018} that have demonstrated FFPs at intensities of $10^{14}$ W/cm$^2$, durations of tens of picoseconds, and spot sizes of $\sigma_0 \sim 10 \lambda_0$ along with rapid developments in laser technology indicate that an experimental demonstration will be realizable in the near term. In fact, intensities beyond the relativistic threshold ($\xi_0 = 1$) are already envisaged for other applications \cite{Palastro_2020}. 

The technology to place an electron beam within the several micron volume of the focus already exists and is regularly used in experiments \cite{powers2014quasi,yan2017high,wu2021high}. The ponderomotive force expelling an off-axis electron from the FFP can be mitigated by starting with higher $\gamma_0$ or by filtering electrons to create a highly collimated beam.

Thus, our present results motivate the forthcoming experimental implementation of FFPs in applications aiming at measuring the dynamics driven by RR, which is to this day a contentious topic, with initial laser-based experiments not yet providing a statistically conclusive observation of RR \cite{Cole_2018,Poder_2018}.

\begin{acknowledgments}
We would like to thank Dustin Froula, Warren Mori, Matteo Tamburini, Jorge Vieira, and Marija Vranic for useful discussions. This material is also based upon work supported by the Office of Fusion Energy Sciences under Award Numbers DE-SC0019135 and DE-SC00215057, the Department of Energy National Nuclear Security Administration under Award Number DE-NA0003856, the University of Rochester, and the New York State Energy Research and Development Authority.
\end{acknowledgments}

%



\section*{Supplementary material}

# Supplemental Material: Radiation Reaction Enhancement in Flying Focus Pulses

M. Formanek,[1] D. Ramsey,[2] J. P. Palastro,[2] and A. Di Piazza[1]

[1] *Max Planck Institute for Nuclear Physics, Saupfercheckweg 1, D-69117 Heidelberg, Germany*
[2] *University of Rochester, Laboratory for Laser Energetics, Rochester, New York, 14623 USA*

The purpose of this supplemental document is to introduce the flying focus fields used in the associated Letter and their properties. We show how to construct flying focus beams and pulses as an exact solution of Maxwell's equations with arbitrary orbital angular momentum $\ell$. Then, we explicitly evaluate the electric and magnetic fields for the case $\ell = 0$ and construct their field invariants. We also present a method of computing the power in the beam going through a transverse plane under the assumption that the phase oscillations of the fields are faster than the rate of change of the intensity envelope along the longitudinal direction. Finally, we describe our implementation of the pulse envelope and the simulation parameters.

## FLYING FOCUS BEAMS AND PULSES

We follow the solution method of Maxwell's equations presented in Ref. [1], which was shown to describe flying focus beams only recently [2], for the case of the focus moving at the speed of light in the opposite direction of the wave fronts. This model is what we use in the Letter and here we extend the solution to arbitrary angular momenta of the beam.

We define light-cone coordinates $\phi = t - z$, $\eta = t + z$ for a wave whose wave fronts move at the speed of light in the direction of the positive $z$-axis (we work in units $\hbar = c = \varepsilon_0 = 1$). In terms of these coordinates, we can write the electromagnetic field four-potential $A^\mu$ in general as

$$A^\mu(\eta, \boldsymbol{r}, \phi) = \frac{1}{2} \int d\omega \widetilde{A}^\mu(\eta, \boldsymbol{r}, \omega) e^{-i\omega\phi} + \text{c.c.}, \tag{1}$$

which is a Fourier transformation in the variables $\omega \to \phi$, and $\boldsymbol{r} = (x, y)$ are the transverse coordinates.

The four-potential components $A_-$ and $A_+$ are defined as $A_- = A^0 - A^z$, $A_+ = A^0 + A^z$. Thus, the Lorenz gauge condition $\partial \cdot A = 0$ in the light-cone coordinates becomes

$$\partial_\phi A_- + \nabla_\perp \cdot \boldsymbol{A}_\perp + \partial_\eta A_+ = 0, \tag{2}$$

where $\partial_\phi = (\partial_0 - \nabla_z)/2$ and $\partial_\eta = (\partial_0 + \nabla_z)/2$. We can set $A_+ = 0$ as additional condition to leave us with two degrees of freedom of the electromagnetic field. By substituting Eq. (1) into Eq. (2), we obtain for the components of $\widetilde{A}^\mu$ the relationship

$$\widetilde{A}_- = -\frac{i}{\omega} \nabla_\perp \cdot \widetilde{\boldsymbol{A}}_\perp. \tag{3}$$

The wave equation $\partial^2 A^\mu = 0$ in light-cone coordinates reads

$$(4\partial_\phi \partial_\eta - \nabla_\perp^2) A^\mu = 0, \tag{4}$$

which gives us in the frequency domain for the $\widetilde{\boldsymbol{A}}_\perp$ components the equation

$$-4i\omega \partial_\eta \widetilde{\boldsymbol{A}}_\perp - \nabla_\perp^2 \widetilde{\boldsymbol{A}}_\perp = 0. \tag{5}$$

This differential equation has exactly the same form of the paraxial equation for the Gaussian beams differing only by a factor of two in the first term. Thus, any solution of this type of equation in the context of Gaussian beams within the paraxial approximation represents here an exact solution of Maxwell's equations in light-cone coordinates. In cylindrical coordinates $(r, \theta, \eta)$, with $r = \sqrt{x^2 + y^2}$ being the distance from the $z$-axis, we can write

$$\left( \partial_r^2 + \frac{1}{r}\partial_r + \frac{1}{r^2}\partial_\theta^2 + 4i\omega\partial_\eta \right) \widetilde{\boldsymbol{A}}_\perp = 0, \tag{6}$$

which has the analytical solution [3]

$$\widetilde{\boldsymbol{A}}_{n,\ell,\perp}(\eta, r, \theta, \omega) = \boldsymbol{\mathcal{A}}_0 f(\omega) \frac{\sigma_0}{\sigma[\eta, \eta_0(\omega)]} \left\{ \frac{\sqrt{2}r}{\sigma[\eta, \eta_0(\omega)]} \right\}^\ell L_n^\ell \left\{ \frac{2r^2}{\sigma^2[\eta, \eta_0(\omega)]} \right\} e^{-r^2/\sigma^2[\eta, \eta_0(\omega)]}$$
$$\times \exp\left\{ i \frac{r^2}{\sigma^2[\eta, \eta_0(\omega)]} \frac{\eta}{\eta_0(\omega)} - i\ell\theta - i(2n + \ell + 1)\arctan\left[\frac{\eta}{\eta_0(\omega)}\right] \right\}, \tag{7}$$

for arbitrary integer $n$ and real $\ell$, where $\sigma[\eta, \eta_0(\omega)]$ relates to the transversal spot size

$$\sigma[\eta, \eta_0(\omega)] = \sigma_0 \sqrt{1 + \eta^2/\eta_0(\omega)^2}, \quad \eta_0(\omega) = \omega \sigma_0^2 \tag{8}$$

and $f(\omega)$ is an arbitrary function which specifies the frequency spectrum. Notice that the Rayleigh length equivalent $\eta_0$ differs by a factor of two from the Gaussian beam Rayleigh length $z_0 = \omega \sigma_0^2/2$. This is a consequence of the factor of two difference in the first term of Eq. (5). At equal spot sizes $\sigma_0$, the Gaussian beams are more confined in the longitudinal direction than the flying focus solutions. Finally, $L_n^\ell$ are the associated Laguerre polynomials

$$L_0^\ell(x) = 1, \tag{9}$$
$$L_1^\ell(x) = 1 + \ell - x, \tag{10}$$
$$L_n^\ell(x) = \frac{x^{-\ell} e^x}{n!} \frac{d^n}{dx^n}(e^{-x} x^{n+\ell}). \tag{11}$$

Since increasing $n$ generates higher order polynomials, we can restrict ourselves to the simplest $n = 0$ case and consider only the $\ell$-dependent solution

$$\widetilde{\boldsymbol{A}}_{\ell,\perp}(\eta, r, \theta, \omega) = \boldsymbol{\mathcal{A}}_0 f(\omega) \frac{\sigma_0}{\sigma[\eta, \eta_0(\omega)]} \left\{ \frac{\sqrt{2}r}{\sigma[\eta, \eta_0(\omega)]} \right\}^\ell e^{-r^2/\sigma^2[\eta,\eta_0(\omega)]}$$
$$\times \exp\left\{ i \frac{r^2}{\sigma^2[\eta, \eta_0(\omega)]} \frac{\eta}{\eta_0(\omega)} - i\ell\theta - i(\ell+1)\arctan\left[\frac{\eta}{\eta_0(\omega)}\right] \right\}. \tag{12}$$

By substituting this solution back into Eq. (1), we obtain the exact solution of Maxwell's equations

$$\boldsymbol{A}_{\ell,\perp}(\eta, r, \theta, \phi) = \frac{1}{2}\boldsymbol{\mathcal{A}}_0 \int d\omega f(\omega) e^{-i\omega\phi} \frac{\sigma_0}{\sigma[\eta, \eta_0(\omega)]} \left\{ \frac{\sqrt{2}r}{\sigma[\eta, \eta_0(\omega)]} \right\}^\ell e^{-r^2/\sigma^2[\eta,\eta_0(\omega)]}$$
$$\times \exp\left\{ i \frac{r^2}{\sigma^2[\eta, \eta_0(\omega)]} \frac{\eta}{\eta_0(\omega)} - i\ell\theta - i(\ell+1)\arctan\left[\frac{\eta}{\eta_0(\omega)}\right] \right\} + \text{c.c.} \tag{13}$$

This solution allows us to construct spatio-temporally localized pulses of finite energy by tuning the spectral profile $f(\omega)$. For example, for the Gaussian frequency profile $f(\omega) = (\sqrt{\pi}/\tau)\exp[-(\omega-\omega_0)^2\tau^2]$ and for $\omega_0\tau \gg 1$, we obtain approximately a Gaussian pulse profile proportional to $\exp(-\phi^2/\tau^2)$ (see [2]). In the limit $\tau \to \infty$ the frequency distribution becomes a delta function

$$f(\omega) = \delta(\omega - \omega_0), \tag{14}$$

which describes a simple monochromatic beam solution with infinite extent in both time and space. For this frequency distribution, the integration can be carried out explicitly just by replacing $\omega \to \omega_0$

$$\boldsymbol{A}_{\ell,\perp}(\eta, r, \theta, \phi) = \frac{1}{2}\boldsymbol{\mathcal{A}}_0 \frac{\sigma_0}{\sigma[\eta, \eta_0(\omega_0)]} \left\{ \frac{\sqrt{2}r}{\sigma[\eta, \eta_0(\omega_0)]} \right\}^\ell e^{-r^2/\sigma^2[\eta,\eta_0(\omega_0)]}$$
$$\times \exp\left\{ -i\omega_0\phi + i\frac{r^2}{\sigma^2[\eta, \eta_0(\omega_0)]} \frac{\eta}{\eta_0(\omega_0)} - i\ell\theta - i(\ell+1)\arctan\left[\frac{\eta}{\eta_0(\omega)}\right] \right\} + \text{c.c.}, \tag{15}$$

which can be evaluated in terms of real functions as

$$\boldsymbol{A}_{\ell,\perp}(\eta, r, \theta, \phi) = \boldsymbol{\mathcal{A}}_0 \frac{\sigma_0}{\sigma[\eta, \eta_0(\omega_0)]} \left\{ \frac{\sqrt{2}r}{\sigma[\eta, \eta_0(\omega_0)]} \right\}^\ell e^{-r^2/\sigma^2[\eta,\eta_0(\omega_0)]}$$
$$\times \cos\left\{ \omega_0\phi - \frac{r^2}{\sigma^2[\eta, \eta_0(\omega_0)]} \frac{\eta}{\eta_0(\omega_0)} + \ell\theta + (\ell+1)\arctan\left[\frac{\eta}{\eta_0(\omega)}\right] \right\}. \tag{16}$$

From Eq. (2) we have the following expression for $A_{\ell,-}$

$$A_{\ell,-}(\eta, r, \theta, \phi) = -\int d\phi \nabla_\perp \cdot \boldsymbol{A}_{\ell,\perp}(\eta, r, \theta, \phi). \tag{17}$$

If we assume that the field is linearly polarized along the $x$-axis, i.e., for $\mathcal{A}_0 = \mathcal{A}_0 \hat{\boldsymbol{x}}$, we can evaluate

$$
\begin{aligned}
A_{\ell,-}(\eta, r, \theta, \phi) = &-\ell \frac{\mathcal{A}_0}{\omega_0} \frac{1}{r} \frac{\sigma_0}{\sigma[\eta, \eta_0(\omega_0)]} \left\{ \frac{\sqrt{2} r}{\sigma[\eta, \eta_0(\omega_0)]} \right\}^\ell e^{-r^2/\sigma^2[\eta, \eta_0(\omega_0)]} \sin[\Psi_\ell(1,0)] \\
&+ \frac{\mathcal{A}_0}{\omega_0} \frac{x}{\sigma^2[\eta, \eta_0(\omega_0)]} \left\{ \frac{\sqrt{2} r}{\sigma[\eta, \eta_0(\omega_0)]} \right\}^\ell e^{-r^2/\sigma^2[\eta, \eta_0(\omega_0)]} \sin[\Psi_\ell(0,1)],
\end{aligned}
\tag{18}
$$

where the first term vanishes for $\ell = 0$. The phases which appear in the above solution are defined as

$$
\Psi_\ell(a, b) = \omega_0 \phi - \frac{r^2}{\sigma^2[\eta, \eta_0(\omega_0)]} \frac{\eta}{\eta_0(\omega)} + (\ell - a)\theta + (\ell + 1 + b) \arctan\left[\frac{\eta}{\eta_0(\omega)}\right].
\tag{19}
$$

Since $A_{\ell,+} = A_\ell^0 + A_\ell^z = 0$ and $A_{\ell,-} = A_\ell^0 - A_\ell^z$ we can evaluate the $A_\ell^0$ and $A_\ell^z$ components as

$$
A_\ell^0 = -A_\ell^z = \frac{1}{2} A_{\ell,-}.
\tag{20}
$$

Note that for any $\ell > 1$, the four-potential on the $z$-axis ($r = 0$) vanishes. The case of $\ell = 1$ is special, because both electric and magnetic fields have only $z$-component non-zero on the axis. Thus, for $\ell = 1$ and in the case of an electron entering such field along the $z$-axis, the motion becomes effectively one-dimensional and radiation reaction has very little effect on the electron trajectory (at least according to the Landau-Lifshitz equation).

In the following we will restrict our analysis on $\ell = 0$, discussed in the Letter, and simplify the notation. For brevity we will denote

$$
\sigma[\eta, \eta_0(\omega_0)] \to \sigma_\eta, \quad \eta_0(\omega_0) \to \eta_0, \quad \Psi_0(0, a) \to \Psi(a),
\tag{21}
$$

i.e., we do not indicate explicitly the dependence on $\omega_0$, and we drop the index $\ell = 0$. Note that in the Letter $\Psi(a)$ is called $\Psi(a, \eta, \eta_0)$ and $\sigma_\eta$ is called $\sigma(\eta, \eta_0)$ to emphasize the dependence on $\eta$, $\eta_0$ when comparing with Gaussian beams.

### ELECTRIC AND MAGNETIC FIELDS ($\ell = 0$)

The electric and magnetic fields can be calculated from the four-vector potential components as

$$
\boldsymbol{E} = -\nabla A^0 - \frac{\partial \boldsymbol{A}}{\partial t}, \quad \boldsymbol{B} = \nabla \times \boldsymbol{A},
\tag{22}
$$

which can be evaluated explicitly knowing the solutions in Eqs. (16) and (18) with $\ell = 0$:

$$
E_x = \mathcal{A}_0 e^{-r^2/\sigma_\eta^2}(T_1 + T_2), \qquad B_x = \frac{2\mathcal{A}_0 xy}{\omega_0 \sigma_0 \sigma_\eta^3} e^{-r^2/\sigma_\eta^2} \sin[\Psi(2)],
\tag{23a}
$$

$$
E_y = \frac{2\mathcal{A}_0 xy}{\omega_0 \sigma_0 \sigma_\eta^3} e^{-r^2/\sigma_\eta^2} \sin[\Psi(2)], \qquad B_y = \mathcal{A}_0 e^{-r^2/\sigma_\eta^2}(T_1 - T_2),
\tag{23b}
$$

$$
E_z = \frac{2\mathcal{A}_0 x}{\sigma_\eta^2} e^{-r^2/\sigma_\eta^2} \cos[\Psi(1)], \qquad B_z = \frac{2\mathcal{A}_0 y}{\sigma_\eta^2} e^{-r^2/\sigma_\eta^2} \cos[\Psi(1)].
\tag{23c}
$$

The two terms $T_1$ and $T_2$ are given by

$$
T_1 = \frac{\omega_0 \sigma_0}{\sigma_\eta} \sin[\Psi(0)],
\tag{24}
$$

$$
T_2 = -\frac{r^2}{\omega_0 \sigma_0 \sigma_\eta^3} \left\{ \frac{2\sigma_0^2 \eta/\eta_0}{\sigma_\eta^2} \cos[\Psi(0)] + \left(1 - \frac{2\sigma_0^2 \eta^2/\eta_0^2}{\sigma_\eta^2}\right) \sin[\Psi(0)] \right\} + \frac{2x^2}{\omega_0 \sigma_0 \sigma_\eta^3} \sin[\Psi(2)].
\tag{25}
$$

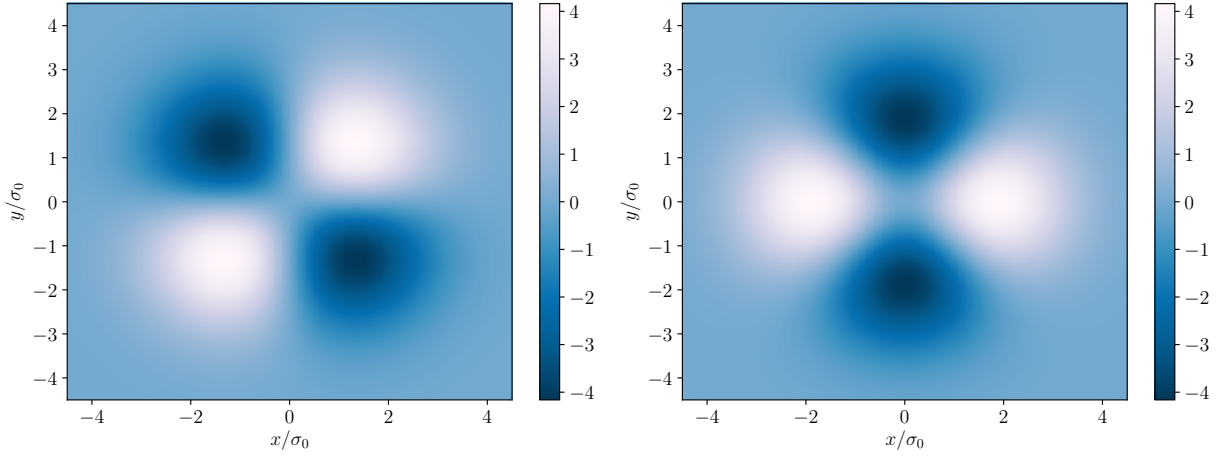

FIG. 1: The invariants $\boldsymbol{E} \cdot \boldsymbol{B}$ (left) and $(E^2 - B^2)/2$ (right) at the focus ($\eta = 0$) for $\sigma_0 = 5$ in arbitrary units. The field is polarized along the $x$-direction.

We can also explicitly construct the field invariants. The invariant $\boldsymbol{E} \cdot \boldsymbol{B}$ is given by

$$\boldsymbol{E} \cdot \boldsymbol{B} = \frac{4\mathcal{A}_0^2 \sigma_0^2}{\sigma_\eta^6} xy e^{-2r^2/\sigma_\eta^2} = \frac{2\mathcal{A}_0^2 \sigma_0^2}{\sigma_\eta^6} r^2 \sin(2\theta) e^{-2r^2/\sigma_\eta^2}, \tag{26}$$

which vanishes whenever either $x$ or $y$ is zero. Similarly, for the invariant $(E^2 - B^2)/2$, we have

$$\frac{1}{2}(E^2 - B^2) = \frac{2\mathcal{A}_0^2 \sigma_0^2}{\sigma_\eta^6}(x^2 - y^2) e^{-2r^2/\sigma_\eta^2} = \frac{2\mathcal{A}_0^2 \sigma_0^2}{\sigma_\eta^6} r^2 \cos(2\theta) e^{-2r^2/\sigma_\eta^2}, \tag{27}$$

which is zero whenever $|x| = |y|$. At the position of the focus $\eta = 0$ we have $\sigma(\eta = 0) = \sigma_0$, and both invariants no longer depend on $z$ and $t$. In Fig. 1 we plot these invariants at focus. Out of the focus $\eta \neq 0$ for either $\eta > 0$ or $\eta < 0$), both invariants look qualitatively the same as at the focus, and only the spot size $\sigma_\eta$ increases as the beam defocuses.

## TIME AVERAGES OF THE OSCILLATING FUNCTIONS

The oscillatory parts of the fields obtained above are given by the trigonometric functions of the type

$$\cos[\Psi(a)] = \cos\left[\omega_0 \phi - \frac{r^2}{\sigma_\eta^2} \frac{\eta}{\eta_0} + (1+a) \arctan\left(\frac{\eta}{\eta_0}\right)\right], \tag{28a}$$

$$\sin[\Psi(a)] = \sin\left[\omega_0 \phi - \frac{r^2}{\sigma_\eta^2} \frac{\eta}{\eta_0} + (1+a) \arctan\left(\frac{\eta}{\eta_0}\right)\right]. \tag{28b}$$

We want to examine how products of these functions behave at some fixed $z$ and perform the time average under the assumption that the oscillations in $\phi$ are much faster than how fast the intensity envelope in $\eta$ varies, i.e., $\sigma_0^2 \gg \lambda_0^2/\pi$. By using the angle addition formulas and the averages

$$\langle \cos^2(\omega_0 \phi) \rangle_z = \frac{1}{2}, \quad \langle \sin^2(\omega_0 \phi) \rangle_z = \frac{1}{2}, \quad \langle \sin(\omega_0 \phi) \cos(\omega_0 \phi) \rangle_z = 0, \tag{29}$$

over time at some fixed $z$, the time-averages of the products of interest then become

$$\langle \sin[\Psi(a)] \cos[\Psi(b)] \rangle_z = \frac{1}{2} \sin\left[(a-b) \arctan\left(\frac{\eta}{\eta_0}\right)\right], \tag{30a}$$

$$\langle \sin[\Psi(a)] \sin[\Psi(b)] \rangle_z = \frac{1}{2} \cos\left[(a-b) \arctan\left(\frac{\eta}{\eta_0}\right)\right], \tag{30b}$$

$$\langle \cos[\Psi(a)] \cos[\Psi(b)] \rangle_z = \frac{1}{2} \cos\left[(a-b) \arctan\left(\frac{\eta}{\eta_0}\right)\right]. \tag{30c}$$

As it is clear from these expressions, the slowly-varying time dependence in $\eta = t + z$ is still kept. Note that the final formulas can be evaluated in terms of $\sigma_\eta$, $\sigma_0$, $\eta$, $\eta_0$ as

$$\sin\left[\arctan\left(\frac{\eta}{\eta_0}\right)\right] = \frac{\sigma_0 \eta/\eta_0}{\sigma_\eta}, \quad \cos\left[\arctan\left(\frac{\eta}{\eta_0}\right)\right] = \frac{\sigma_0}{\sigma_\eta}, \tag{31a}$$

$$\sin\left[2\arctan\left(\frac{\eta}{\eta_0}\right)\right] = 2\frac{\sigma_0^2}{\sigma_\eta^2}\frac{\eta}{\eta_0}, \quad \cos\left[2\arctan\left(\frac{\eta}{\eta_0}\right)\right] = \frac{\sigma_0^2}{\sigma_\eta^2}\left(1 - \frac{\eta^2}{\eta_0^2}\right), \tag{31b}$$

which can be continued for higher $(a-b)$ values using the angle addition formulas.

## POWER IN THE FLYING FOCUS BEAM

The average power flowing through the $xy$ plane at a given $z$ can be calculated as the Poynting vector flux

$$P_{\text{ave}} = \int dxdy \langle \boldsymbol{E} \times \boldsymbol{B}\rangle_z \cdot \hat{\boldsymbol{z}}, \tag{32}$$

where $\hat{\boldsymbol{z}}$ is the unit vector normal to the $xy$ plane, i.e., $\hat{\boldsymbol{z}} = (0, 0, 1)$. Therefore, we need to evaluate the quantity

$$P_{\text{ave}} = \int dxdy \langle E_x B_y - E_y B_x\rangle_z. \tag{33}$$

The second term is easily computed as

$$\int dxdy \langle E_y B_x \rangle_z = \int dxdy \frac{4\mathcal{A}_0^2 x^2 y^2}{\omega_0^2 \sigma_0^2 \sigma_\eta^6} e^{-2r^2/\sigma_\eta^2} \langle \sin^2[\Psi(2)]\rangle_z = \frac{\pi}{16\omega_0^2 \sigma_0^2}\mathcal{A}_0^2. \tag{34}$$

The first term reads

$$\int dxdy \langle E_x B_y \rangle_z = \int dxdy \mathcal{A}_0^2 e^{-2r^2/\sigma_\eta^2} \langle T_1^2 - T_2^2 \rangle_z. \tag{35}$$

The first part of this expression proportional to $T_1^2$ can be again directly evaluated, and the result is

$$\int dxdy \mathcal{A}_0^2 e^{-2r^2/\sigma_\eta^2} \langle T_1^2 \rangle_z = \int dxdy \mathcal{A}_0^2 e^{-2r^2/\sigma_\eta^2} \frac{\omega_0^2 \sigma_0^2}{\sigma_\eta^2} \langle \sin^2[\Psi(0)]\rangle_z = \frac{\pi}{4}\mathcal{A}_0^2 \omega_0^2 \sigma_0^2. \tag{36}$$

Finally, the last term is given by

$$\int dxdy \mathcal{A}_0^2 e^{-2r^2/\sigma_\eta^2} \langle T_2^2 \rangle_z \tag{37}$$

and we need to evaluate the square and the time average of the expression (25). After some algebraic manipulations and using the formulas (30a-30c) for the time averages, we obtain

$$\langle T_2^2 \rangle_z = \frac{2}{\omega_0^2 \sigma_\eta^6 \sigma_0^2}\left(r^4/4 + x^4 - r^2 x^2\right). \tag{38}$$

Now, we can integrate this expression over the transversal plane:

$$\int \mathcal{A}_0^2 e^{-2r^2/\sigma_\eta^2}\langle T_2^2\rangle_z dxdy = \frac{\pi}{16\omega_0^2 \sigma_0^2}\mathcal{A}_0^2. \tag{39}$$

We see that in all of the integrals above the dependence on $\eta$ drops out, as one would expect, and the result is constant and finite. Altogether for the power we have

$$P_{\text{ave}} = \frac{\pi}{4}\mathcal{A}_0^2\left(\omega_0^2\sigma_0^2 - \frac{1}{2\omega_0^2\sigma_0^2}\right) \approx \frac{\pi}{4}\mathcal{A}_0^2\omega_0^2\sigma_0^2 \doteq 21.5[\text{GW}]\left(\xi_0\frac{\sigma_0}{\lambda_0}\right)^2, \tag{40}$$

where $\xi_0 = |e|\mathcal{A}_0/m$ is the dimensionless normalized field amplitude. Consistently with our approximations, the second term in Eq. (40) can be neglected. In fact, already for $\sigma_0 = \lambda_0$ the error resulting from omitting this term is $\approx 0.03\%$. The full expression shows that with decreasing $\sigma_0$, the power decreases until it would vanish at $\sigma_0/\lambda_0 = 2^{-5/4}\pi^{-1} \approx 0.14$. However, for such small value of $\sigma_0$, our assumption that the field is oscillating much faster than the envelope no longer applies, and the result for the average beam power is not valid. Finally, we would like to point out that the resulting expression is the same as the equivalent expression for the Gaussian beams, with $\sigma_0$ being the spot size at focus (see for example [4, 5]).

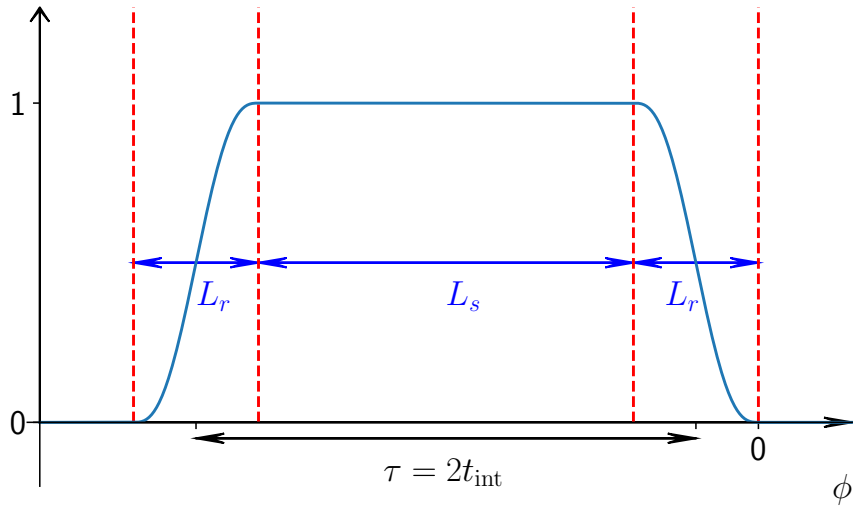

FIG. 2: The $g(\phi)$ pulse envelope function. $L_r$ is the length of the ramps and $L_s$ the length of the flattop region.

## PULSE ENVELOPE

As we described in the Letter we implemented the pulse envelope $g(\phi)$ on the level of a multiplicative factor modifying the fields (23a-23c). In doing so, the fields no longer satisfy the Maxwell equations exactly, but as long as the derivatives of the function $g(\phi)$ are small compared to the derivatives of the fields itself, the error is not large and is confined to ramp on/off regions. The functional prescription of the function $g(\phi)$ is given by 5th order polynomials ensuring smooth transition between the regions where $g(\phi) = 0$ and $g(\phi) = 1$ including the first derivatives. Specifically, we have

$$g(\phi) = \begin{cases} -10\left(\frac{\phi}{L_r}\right)^3 - 15\left(\frac{\phi}{L_r}\right)^4 - 6\left(\frac{\phi}{L_r}\right)^5 & -L_r < \phi \leq 0\,, \\ 1 & -L_r - L_s < \phi \leq -L_r\,, \\ 10\left(\frac{\phi+2L_r+L_s}{L_r}\right)^3 - 15\left(\frac{\phi+2L_r+L_s}{L_r}\right)^4 + 6\left(\frac{\phi+2L_r+L_s}{L_r}\right)^5 & -2L_r - L_s \leq \phi \leq -L_r - L_s\,, \\ 0 & \text{otherwise}\,, \end{cases} \quad (41)$$

where $L_r$ indicates the length of the ramps and $L_s$ is the length of the flattop region. We set the length of the pulse $\tau$ to be equal to $\tau = L_s + L_r$ to minimize the effect of the ramps. We refer to Fig. 2 for illustration. The whole pulse envelope can be also displaced by replacing $\phi \to \phi + \phi_0$ ensuring the correct timing of the electron interaction with Gaussian/flying focus pulses. Finally, we chose the length of the ramp $L_r$ to be equal to five Rayleigh ranges ($5\eta_0$ for FFPs) consistently for all simulations. This length is arbitrary as long as it is not too short (to violate the slowly-changing pulse envelope approximation) or too long (to significantly affect the pulse length). We compare the simulations with analytical estimates assuming rectangular pulse approximation and the results are robust with respect to ramp length variations.

## SIMULATION PARAMETERS

In the numerical simulation examples discussed in the Letter, the flying focus pulses were kept at total energy $E_{\text{tot}} = 10$, 50, and 200 J. Since the laser wavelength was set to $\lambda_0 = 1$ $\mu$m the interaction time of 100 ps corresponds to $t_{\text{int}} = 188,400$ $1/\omega_0$ in laser units. This means that the average power is $P_{\text{ave}} = 0.05$, 0.25, and 1 TW respectively. The time of the free propagation of the electron before encountering the envelope $g(\phi)$ was set to $t_{\text{free}} = 7.5$ $1/\omega_0$. The integration step was chosen to be $dt = 0.01$ $1/\omega_0$ which gives us about 300 steps per one laser wavelength. All other simulation parameters which are function of the spot size $\sigma_0$ are summarized in the Table I.

In our simulation examples the pulse length $\tau = L_s + L_r = 2t_{\text{int}} \approx 376,800 1/\omega_0$ which is many times longer than the Rayleigh range. The largest Rayleigh range we consider is for $\sigma_0/\lambda_0 = 8$ when the envelope is stil about 150 times

| | pulse geometry | | | $\xi_0$ | | |
|---|---|---|---|---|---|---|
| $\sigma_0/\lambda_0$ | $\eta_0(1/\omega_0)$ | $L_s(10^5\,1/\omega_0)$ | $L_r(1/\omega_0)$ | $E_\text{tot}=10$ J | $E_\text{tot}=50$ J | $E_\text{tot}=200$ J |
| 2.50 | 246.7 | 3.756 | 1234 | 0.6100 | 1.364 | 2.728 |
| 2.75 | 298.6 | 3.753 | 1493 | 0.5545 | 1.240 | 2.480 |
| 3.00 | 355.3 | 3.750 | 1777 | 0.5083 | 1.137 | 2.273 |
| 3.25 | 417.0 | 3.747 | 2085 | 0.4692 | 1.049 | 2.098 |
| 3.50 | 483.6 | 3.744 | 2418 | 0.4357 | 0.9743 | 1.949 |
| 3.75 | 555.2 | 3.740 | 2776 | 0.4067 | 0.9093 | 1.819 |
| 4.00 | 631.7 | 3.736 | 3158 | 0.3812 | 0.8525 | 1.705 |
| 4.25 | 713.1 | 3.732 | 3565 | 0.3588 | 0.8023 | 1.605 |
| 4.50 | 799.4 | 3.728 | 3997 | 0.3389 | 0.7578 | 1.516 |
| 4.75 | 890.7 | 3.723 | 4454 | 0.3210 | 0.7179 | 1.436 |
| 5.00 | 987.0 | 3.719 | 4935 | 0.3050 | 0.6820 | 1.364 |
| 5.25 | 1088 | 3.714 | 5441 | 0.2905 | 0.6495 | 1.299 |
| 5.50 | 1194 | 3.708 | 5971 | 0.2773 | 0.6200 | 1.240 |
| 5.75 | 1305 | 3.703 | 6526 | 0.2652 | 0.5930 | 1.186 |
| 6.00 | 1421 | 3.697 | 7106 | 0.2542 | 0.5683 | 1.137 |
| 6.25 | 1542 | 3.691 | 7711 | 0.2440 | 0.5456 | 1.091 |
| 6.50 | 1668 | 3.685 | 8340 | 0.2346 | 0.5246 | 1.049 |
| 6.75 | 1799 | 3.678 | 8994 | 0.2259 | 0.5052 | 1.010 |
| 7.00 | 1934 | 3.671 | 9672 | 0.2179 | 0.4871 | 0.9743 |
| 7.25 | 2075 | 3.664 | 10380 | 0.2103 | 0.4703 | 0.9407 |
| 7.50 | 2221 | 3.657 | 11100 | 0.2033 | 0.4547 | 0.9093 |
| 7.75 | 2371 | 3.649 | 11860 | 0.1968 | 0.4400 | 0.8800 |
| 8.00 | 2527 | 3.642 | 12630 | 0.1906 | 0.4262 | 0.8525 |

TABLE I: FFP simulation parameters for $E_\text{tot} =$ 10, 50 and 200 J and interaction time $t_\text{int} = 100$ ps $\approx 188,400\,1/\omega_0$

longer. As indicated above, the pulse ramps vary on the Rayleigh range scale as well and at most entail $\sim 3\%$ of the total pulse length. This justifies the use of rectangular pulse approximation in our analytical estimates.

From the Table I we see that the requirements on the pulse amplitude $\xi_0$ with given parameters lie between 0.2 - 2.7, intensities which are already envisaged for flying focus pulses [6]. From the spot sizes in the first column of Table I the laser's $f-$numbers can be calculated to lie in range $f^\# \in (4, 12.5)$, avoiding micro-focusing regime.

---


\end{document}